# Multiphase Flow Simulation of Blow-by and Fuel-in-Oil Dilution via the Piston Ring Pack Using the CFD Level-Set Method

Patrick Antony[1], Norbert Hosters[1], Marek Behr[1],
Anselm Hopf[2], Frank Krämer[2], Carsten Weber[2], Paul Turner[2]

1: Chair for Computational Analysis of Technical Systems RWTH Aachen University, Schinkelstr. 2, 52062 Aachen, Germany
2: Ford of Europe

## Abstract

Modern diesel engines temporarily use a very late post-injection in the combustion cycle to either generate heat for a diesel particulate filter regeneration or purge a lean NOx trap. In some configurations, unburned fuel is left at the cylinder walls and is transported via the piston rings toward the lower crankcase region, where fuel may dilute the oil. Reduced oil lubrication shortens the oil service intervals and increases friction. Beside diesel fuel, this problem may also occur for other types of liquid fuels such as alcohols and e-fuels. The exact transport mechanism of the unburned fuel via the piston-ring-pack grooves and cylinder wall is hard to measure experimentally, motivating numerical flow simulation in early design stages for an in-depth understanding of the involved processes.

A new CFD simulation methodology has been developed to investigate the transient, compressible, multiphase flow around the piston ring pack, through the gap between piston and liner, and its impact on fuel or oil transport. The modern level-set approach is used for the multiphase physics, which directly captures the sharp interface between blow-by gas and fuel or oil. Transient blow-by and two-phase flow simulations have been extensively applied to a Ford 2.0L I4 diesel test engine. The results confirm the validity of the flow compressibility assumption and highlight the sensitivity of the fuel leakage regarding piston sealing ring movement and highly resolved meshes for the multiphase flow. Based on the simulation results, design recommendations for piston and piston ring geometry are provided to reduce the fuel transport toward the crankcase.

## Introduction

This work aims to understand the transport mechanisms of fuel-in-oil by investigating the gas flow around the piston ring pack, through the gap between piston and liner, and its impact on fuel and oil transport. This is done using computational fluid dynamics (CFD) numerical simulation of the gas flow via the piston ring pack into the crankcase to analyze and understand the transport of oil and unburned fuel to reduce oil dilution. In a first step, the flow of blow-by gases is investigated in a Ford piston ring pack (see Fig. 1) with a two-dimensional axisymmetric model. In a second step, the model is extended with a two-phase flow formulation to capture the combined flow of gas and fuel or engine oil (see Fig. 2).

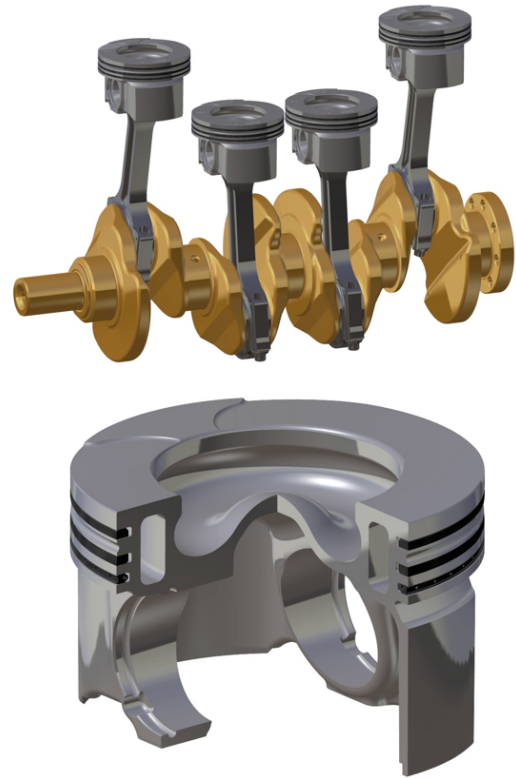

Figure 1: Drive-train of the engine and cut through piston with piston rings.

### State of the Art

The piston ring pack of an internal combustion engine (ICE) is involved in several physical processes and has been investigated regarding multiple optimization goals. One of its main tasks is sealing off the high-pressure combustion chamber from the low-pressure crankcase. Pressure losses due to the so-called blow-by decrease engine efficiency [1]. These have been investigated by experiments and via CFD simulations.

The piston rings also transport heat from the piston to the liner, where the contact to the liner surface is usually lubricated with engine oil to decrease friction forces. Exhaust gas re-circulation (EGR) and incomplete combustion can leave unburned fuel on the cylinder walls. Late post-injection strategies motivated by a reduction of soot in modern particulate filters (as well as increased usage of high-octane biodiesel) increase the amount of unburned fuel in the combustion chamber [2]. This leftover fuel is then transported through the piston ring pack toward the lower crankcase and dilutes the lubrication oil (see Fig. 2). The driving forces causing this transport are blow-by gas, dragging the fuel film toward the crankcase, transported atomized fuel, and the complex scraping and pumping motions of the sealing rings designed to keep the oil away from the combustion chamber. Reverse blow-by and ring scraping on the up-strokes (compression and exhaust) move fuel and oil back toward the combustion chamber. At and below the oil control ring, leaking fuel is very likely to encounter oil, since it is sprayed on the bottom side of the piston for oil-jet cooled pistons. The fuel mixes well with the oil and reduces its lubrication effectiveness, sometimes even reacting with and consuming additives of the oil [2]. Experimental methods on the dilution process are reviewed in [3]. Oil dilution causes increased engine wear and necessitates increased maintenance, either to replace the diluted oil or replace worn out parts.

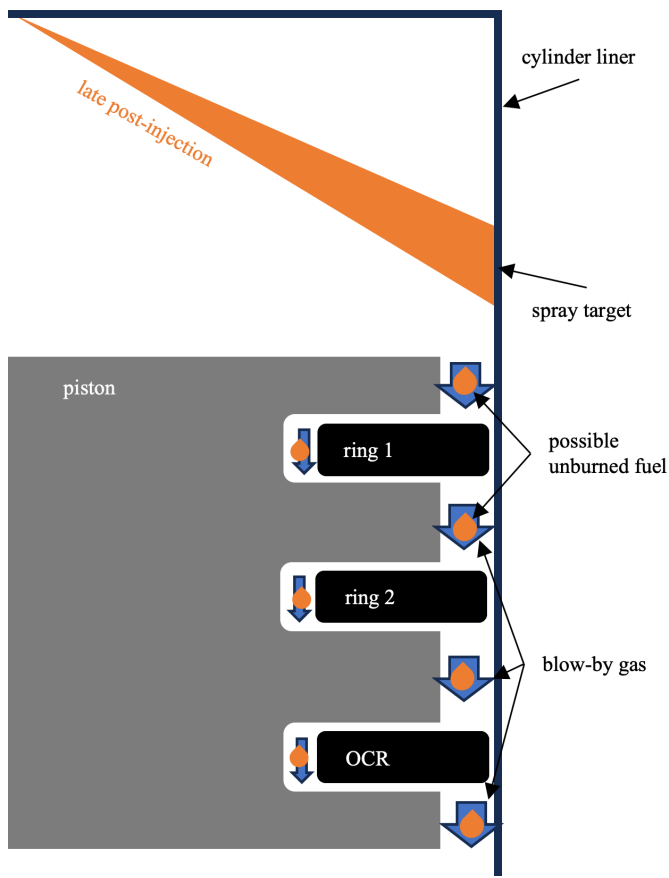

Figure 2 Possible transport path of left-over fuel.

Experiments using fluorescent oil showed oil transport by blow-by gases and ring motion [4], especially at ring flutter, ring collapse, and through the ring end gaps. However, these methods are unable to detect the transport of atomized oil. A follow-on study [5] investigates measures to increase oil and fuel transport toward the crankcase motivated by reduced oil burn off in the combustion chamber. These measures increase the blow-by and fuel-to-oil leaks, which reiterate the complex interactions and design goals of the piston ring pack, where oil burn off in the combustion chamber must be balanced against fuel leakage into the oil.

A similar setup was used to investigate the oil transport from the liner surface to the moving piston land via the piston rings [6]. This effect was also modeled using 2D CFD simulations for different ring and chamfer geometries [7] and taking piston secondary motion (thrust side tilt) [8] into account. The thrust or anti thrust side tilt of the piston induces asymmetry of the piston side clearance and affects ring scraping differently around the piston [9]. Veetil et al. [10] introduce a 3D CFD model of gas blow-by and oil transport with fixed ring positions. Koszałka and Guzik [11] include ring motion and oil film in a simplified ordinary differential equation gas flow model.

A more complete piston dynamics model is presented in [12], predicting ring bending near the ring end gaps and reiterating the significant influence of the ring position within the groove on blow-by. Many of the main oil transport mechanisms depend on piston ring dynamics: Blow-by and reverse blow-by heavily depend on the current ring position within the groove and squeezing during the axial motion of the rings [13]. Similar models were used to derive measures reducing blow-by [14,15]. A 2D blow-by simulation method incorporating axial ring motion with re-meshing of the fluid domain is introduced in [13] and extended with a multiphase model to account for oil transport [16]. These papers use the volume of fluid (VOF) method to model the multiphase physics. The 2D results represent the oil phase with an ambiguous gas-oil interface leading to a reduced accuracy of the oil and air surface curvature and surface tension along the piston ring pack due to the chosen multiphase model.

To the authors' knowledge, no other model currently includes ring motion and multiphase flow effects of gas and fuel or oil with sharp interfaces, for example using the level-set method. Also, fuel entering the crown land using a sharp interface model has not been presented before in literature.

## Problem Description

The fluid flow is investigated around the piston ring pack of a single piston of the Ford 2.0L I4 diesel test engine with four cylinders. The engine specifications are listed in Tab. 1. One engine cycle is decomposed into 720 degrees of crank angle starting at the beginning of the compression stroke. Simulations exceeding one or two engine cycles are currently unfeasible (e.g., a full drive cycle) due to the vast amount of computational power required to resolve the fine time scales of the gas flow. The drive-train of the engine and a cut through the piston is shown in Fig. 1.

We choose a frame of reference moving with the piston. Thus, the only moving walls are the piston rings and the cylinder liner or block wall. The combustion chamber is not resolved, but measured pressure profiles and consistent temperature profiles from 1D simulation are assigned as boundary conditions at the top of the upper crown land. For the 2D cross-section, the oil reservoir toward the piston skirt is included in the modeled geometry (see Fig. 3a). The pressure and temperature within the crankcase are assumed to be constant over the cycle at measured values.

| Fuel | Diesel |
|---|---|
| Cubic capacity | 1995 ccm |
| No. Cylinders | 4 |
| Cylinder block material | Grey cast iron |
| Combustion | Turbo direct common rail injection |
| Max. Power | Up to 140 kW |
| Max. Torque | Up to 420 Nm |
| Stroke | 90 mm |
| Bore | 84 mm |

Table 1: Engine specifications of the Ford 2.0L I4 diesel test engine.

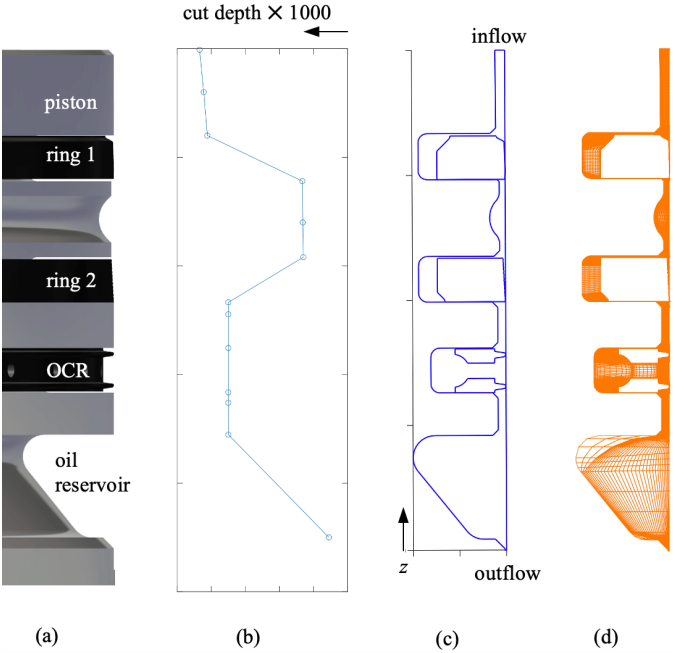

Figure 3: Extracted flow domain of cross-section cut through piston and piston rings. From left to right: (a) CAD raw data (3D, side view); (b) Material cut off from piston lands (not to scale); (c) 2D fluid domain boundaries; (d) Coarse 2D mesh.

## Simulation Method

The simulation was developed with an in-house CFD code based on stabilized finite elements [17,18]. It solves conservation equations for mass, momentum, and energy in the fluid domain, as well as a virtual elasticity problem to track the domain movement. A level-set advection equation is used for multiphase flow problems [19,20] to capture the interface between the two immiscible non-stationary Newtonian fluids oil and air. Furthermore, gravity and surface tension forces must be considered, e.g., to capture the important formation of realistic droplets and possible jet atomization. The solver is parallelized with MPI for HPC computing, has several physical models implemented in an object-oriented way, and solves large sparse linear systems resulting from the Newton-Raphson iterations using a version of GMRES.

### Geometry Modeling

The fluid geometry for the sealing labyrinth around the piston rings was extracted from CAD and engine data. As in [16], the spring inside the oil control ring is neglected and for the 2D cut, the worst-case scenario of the cut is chosen. Thus, the cut passes through the center of one of the holes in the oil control ring (see Fig. 3c). The CAD data represents piston, rings, and liner or cylinder block in a raw machined or forged shape. To model the ring pack labyrinth as close as possible to the actual geometry assembled in the engine, several machining processes must be considered because they substantially change the considered geometry:

The first extraction from the CAD data is a two-dimensional cut through the piston: the extracted profile is assumed to be radially symmetric, but dependent on the z-coordinate in Fig. 3. In a further manufacturing step, the piston is ground. The grinding profile increases the piston's clearance to the liner from 5 to 50μm (see Fig. 3b). Additionally, sharp edges within the CAD geometry impact flow separation and vortices for CFD without pre-processing the geometry. However, they are not present in the assembled piston since they are machined and cut off as well. This results in chamfers as presented in the final geometry, Fig. 3c. Machining tolerances are given for all machining steps mentioned above and introduce uncertainty of a few micrometers to the model. All uncertain cut parameters are assumed at their mean value; thus, no further changes result for the cut profile. Deposition and wear off effects like in [21], as well as thermal expansion effects were ignored in this work.

### Moving Domain

The flow domain is changing during each engine cycle through movement of the piston rings, thermal expansion of the piston, and a possible varying radius of the cylinder bore, the so-called bore distortion. We consider only primary piston ring motion in this work. The ring movement influences the blow-by significantly [21]. To capture the movement of the walls surrounding the fluid domain without re-meshing, the Elastic Mesh Update Method (EMUM [22]) is used to calculate the displacements $d_{\#}$ of the inner nodes of the 'top' (i.e. next time-step) part of the space-time fluid mesh. Given displacements at the surrounding walls $d_{\text{wall}}$, EMUM computes the displacement as solution to a virtual elastic solid equation problem:

$$\nabla \cdot \sigma_{\#}(d_{\#}) = 0$$

$$d_{\#} = d_{\text{wall}}$$

The material parameters of the virtual mesh stress $\sigma_{\#}(d_{\#})$ are chosen such that the mesh retains a high quality after the deformation. In a fluid-structure-contact interaction problem, the displacements $d_{\text{wall}}$ are computed by the structure or contact problem. In this work, $d_{\text{wall}}$ is prescribed from the piston movement profile, fixed at zero, or interpolated between moving and non-moving regions. Since the whole movement of the domain is assumed to be known a priori, any test cases can be run without fluid flow to check the mesh and motion profile for possible deterioration.

The initial fluid domain is meshed in two dimensions using patch-wise structured quadrilateral linear (first order) elements using the transfinite feature in Gmsh [23]. The contact of the piston rings with their respective groove flanks and the contact between the piston ring and cylinder liner or block wall are modeled with the residual gap method. Here, a small gap with few elements and negligible width (approx. 2 − 5μm) remains between the two solid bodies to model the contact.

### Piston Movement with Crank and Pin Offset

The piston ring movement, as well as the liner velocity in a piston-fixed reference frame, is heavily dependent on the piston's primary movement. Modern engine architectures use so-called crank offsets or bore offsets to reduce engine friction. The Ford test engine has a crank or bore offset of −10mm, which also shifts the top dead center (TDC) and bottom dead center (BDC) by some degrees crank angle versus time.

| Engine property | Symbol | Unit | Value |
|---|---|---|---|
| rotations per minute | $N$ | r/min | ~1500 |
| angular velocity | $\omega$ | rad/s | 157.1 |
| stroke length | $s_{max}$ | mm | ~ 90.0 |
| crank offset | $c$ | mm | −10.0 |
| piston pin offset | $p$ | mm | 0.3 |

Table 2: Engine parameters for chosen load case.

Ignoring the piston's secondary movement, the piston position $s$ is calculated according to the formula derived by [24], rewritten with a factor of $2\pi$ included in this work's definition of $\omega = N \cdot \pi/30$ as

$$s(t) = r\cos(\omega t) + \sqrt{b^2 - (r\sin(\omega t) + d)^2}.$$

Here, $r$ is the crank radius, $b$ the con rod length, and $d$ the total offset, which is defined as sum of crank offset $c$ and pin offset $p$, so $d = c + p$ as displayed in Fig. 4.

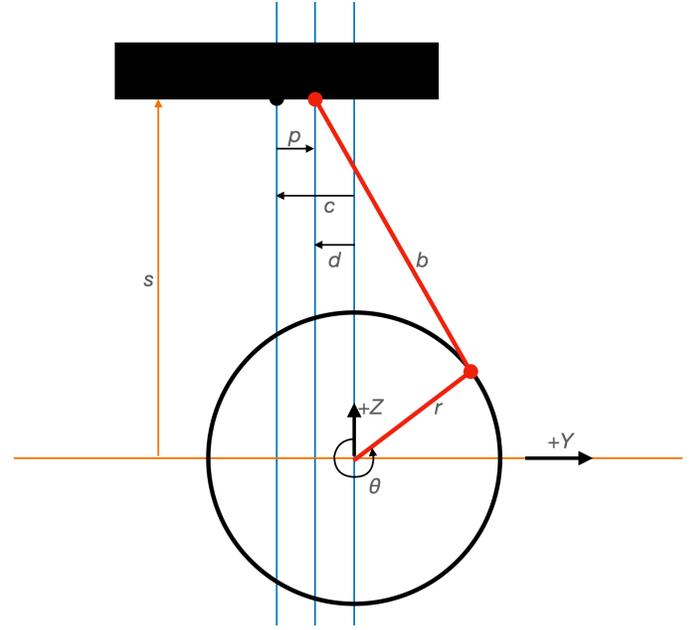

Figure 4: Geometry definitions (not to scale) for piston primary movement: piston position above crank $s$, crank angle $\theta$, crank radius $r$, con rod length $b$, crank offset $c$, pin offset $p$ and total offset $d$.

### Compressible Navier-Stokes Equations

All chemical reactions in the ring pack are assumed to be completed or negligible. The channel width in the piston ring pack is small in the order of 50μm, but due to high pressures (more than atmospheric), non-equilibrium effects are neglected. The liquid phase of the two-phase flow is often modeled as incompressible [25,26]. In this work, we use a material model in the frame of a compressible advective-diffusive system for easy coupling to the compressible gas phase and consistent boundary description. One way of modeling compressible and nearly incompressible phases within such a model is the stiffened ideal gas equation [27]. For now, we restrict the material model to an ideal gas law for both phases and neglect the differences to a stiffened ideal gas model.

Thus, the dynamic behavior of fluids as a continuum is described by a compressible ideal gas model which maps the conservative variables $\boldsymbol{U}$, consisting of density of mass $\rho$, momentum $\rho \boldsymbol{u}$, and energy $\rho e_{tot}$ as

$$\boldsymbol{U} = \begin{pmatrix} \rho \\ \rho \boldsymbol{u} \\ \rho e_{tot} \end{pmatrix} = \begin{pmatrix} p/(RT) \\ \boldsymbol{u}p/(RT) \\ c_v p/(R) \end{pmatrix}$$

to the vector of pressure-primitive variables $\boldsymbol{Y} = (p, \boldsymbol{u}^T, T)^T$, consisting of pressure $p$, velocity vector $\boldsymbol{u}$, and temperature $T$. Here, $e_{tot}$ denotes the internal total energy, $R$ the specific ideal gas constant, and $c_v$ the specific heat at constant volume. The components of the velocity vector $\boldsymbol{u}$ are denoted $u$ and $v$ for two dimensions. The temporal evolution of the conservative variables [17] forms a hyperbolic system of equations

$$\frac{\partial \rho}{\partial t} + \nabla \cdot (\rho \boldsymbol{u}) = 0$$

$$\frac{\partial (\rho \boldsymbol{u})}{\partial t} + \nabla \cdot (\rho \boldsymbol{u} \otimes \boldsymbol{u}) + \nabla p - \nabla \cdot \boldsymbol{\tau} = \boldsymbol{f}$$

$$\frac{\partial (\rho e_{\text{tot}})}{\partial t} + \nabla \cdot (\rho e_{\text{tot}} \boldsymbol{u}) + \nabla \cdot (p \boldsymbol{u}) - \nabla \cdot (\boldsymbol{\tau} \boldsymbol{u}) + \nabla \cdot \boldsymbol{q} = 0$$

with heat flux $\boldsymbol{q}$, stress tensor $\boldsymbol{\tau}$, and volume forces $\boldsymbol{f}$. Dirichlet boundary conditions are applied for velocity on no slip walls, pressure on inflow and outflow, and temperature on all boundaries except the outflow, respectively.

The volume forces $\boldsymbol{f}$ consist of acceleration terms $-\frac{\partial^2 s(t)}{\partial t^2} \rho$ and volumetric surface tension forces $\boldsymbol{f}_{\text{st}}$.

Within this model, flow turbulence is handled as 'residual based LES' according to Rajanna et al. [28]. The initial condition for the very first time-step is computed from a steady-state solution. Details on the implementation can be found in [17].

### Interface Modeling

The complex motion of blow-by gas and liquid oil flow is simulated as a one-fluid model, so both phases share one continuous field of conservative variables. This model requires resolving steep gradients or even discontinuities and an approach of modeling the interface to be able to assign material properties to locations in each respective phase. There are several methods to model the interface, mainly subdivided into interface tracking and interface capturing methods [29].

Interface tracking methods, which compute the position of the (in our case) fluid-fluid interface explicitly, are rarely used for turbulent two-phase flows, as large total displacements of the phase boundary require intensive recalculations of the computational geometry in the form of re-meshing or mesh updates.

More common are approaches from the class of interface capturing, which incorporate information about the phase field implicitly by a function. In this category, one can list the level-set methods, volume of fluid methods, the marker-and-cell method, as well as phase field methods, and combinations of those [30].

The volume of fluid (VOF) method was introduced by Noh and Woodward [31], refined by Hirt and Nichols [32], and used for the two-phase flow analysis in piston rings packs [16]. In this method, a color or fraction function keeps track of the fraction of fluid in each volume cell (see Fig. 5). Cells filled entirely with oil have a fraction of 1, and cells filled entirely with air have a fraction value of 0. Cells containing both oil and air, usually at the interface, have a volume fraction between 0 and 1. The method is inherently mass conserving, but overly diffusive. Often, arbitrary threshold values are used to compute the interface location for VOF methods, leading to inaccuracies of the fluid representation and consequently in the interface normal vectors and curvature too (see Fig. 5). Furthermore, the computation of the surface tension force, which requires the interface curvature, is not as accurate as in the level-set method. Newer versions of VOF methods try to address this problem by reconstructing an approximation of the sharp interface location in a second step after advection of the mass fractions.

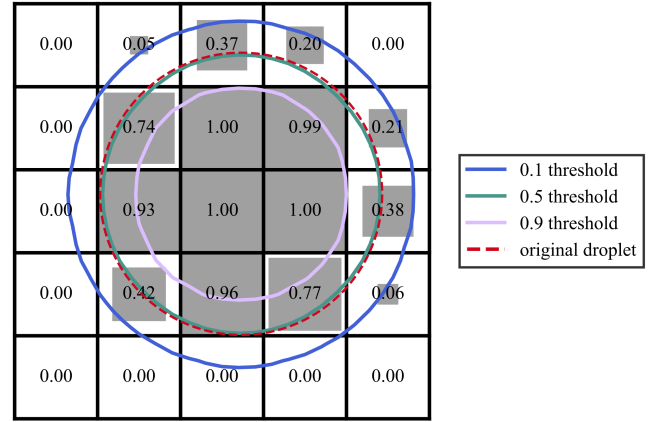

Figure 5: Extracted droplet representation from a smoothed interface for a VOF method with three threshold levels (0.9, 0.5, and 0.1) overlaid on a mesh. The smoothing mimics diffusion of the color function at the advection step. The threshold levels are highlighted to get a sense of the differences introduced in the droplet surface by an arbitrary threshold value used in the VOF method. In this example, the unique solution of the level-set method corresponds to the shape of the original droplet.

### Level-Set Method

In this work, a level-set method with continuous finite elements is used. The level-set method in combination with standard linear space-time finite elements was successfully employed for oil sprays [33]. In this work, discontinuities are smeared out over few element lengths to avoid oscillations like in the treatment of shocks in the compressible solver. Surface tension is included via the continuous surface force approach introduced by [34] as implemented by [33]. The level-set method used in this work is modified from [19,35] for compressible flow. In this work, the level-set function $\phi$ is a signed distance function – thus, the interface is represented by its zero-iso-contour set. The function is initialized as the signed distance to the closest point on the interface and is advected as

$$\frac{\partial \phi}{\partial t} + \nabla \cdot (\phi \boldsymbol{u}) = 0$$

in an iteratively staggered scheme after each Newton iteration of the compressible Navier-Stokes solver. The signed distance property is lost through the advection step but is restored with a re-distancing step in a narrow band around the interface.

From the level-set, the material properties are interpolated between the two phases as in [36] with a smoothing function

$$f(x) = f_1(x) + ((f_2(x) - f_1(x)) \left( \frac{1}{2} + \frac{\phi}{2\epsilon} + \frac{\sin\left(\frac{\pi \phi}{\epsilon}\right)}{2\pi} \right),$$

where $f_1(x)$ and $f_2(x)$ are the material properties of the two phases respectively, and $\epsilon$ is the numerical interface thickness. Here, $\epsilon$ is set to 3.125 mesh element lengths to avoid discontinuities in the solution.

Advection and re-distancing introduces mass errors in the captured phase, which are corrected globally by a scheme introduced in [37] with two modifications: Instead of a correction scheme based on the volume, the mass of the liquid phase is corrected. Secondly, interface smoothing (see the equation above with $f_1 = \rho$ and $f_2 = 0$) is incorporated into the mass computation, instead of using a sharp cutoff at the zero-level set.

## Liquid Phase

For the investigation of the two-phase flow, the liquid phase in the piston ring pack is either unburnt fuel, leftover from combustion or sprayed onto the cylinder wall during late post-injection, or engine oil sprayed up to the piston skirt or lower part from the oil cooling jet or scraped from the oil-wetted cylinder liner or block wall. Relevant material properties for diesel fuels and engine oils are listed in Tab. 3.

| Medium | $\rho$ [kg/m³] | $\mu \times 10^{-3}$ [kg/m·s] | $\gamma$ [kg/s²] |
|---|---|---|---|
| Diesel @20°C [38] | 834.7 | 3.53 | - |
| Diesel @40°C [39] | 810−860 | 1.215−5.16 | - |
| Diesel @60°C [40] | 804.3 | - | 0.0248 |
| Biodiesel @60°C [40] | 851.3 | - | 0.028 |
| 5W-30 Oil @60°C [41] | 833 | 25.38 | 0.03030 |
| Air @60°C | 1.136 | 0.01992 | NA |

Table 3: Properties of oil and diesel fuel: density $\rho$, dynamic viscosity $\mu$, and surface tension to air $\gamma$.

Both 'diesel (fuel)' and 'engine oil' are rather category names than specific substance compositions and their properties can therefore vary depending on the manufacturer and sample. Different fuel types, like winter or summer diesel, and biodiesel fuels with different rates of canola oil content have different material properties. Moreover, properties such as density and viscosity are highly dependent on temperature [38]. From reviewed literature (see Tab. 3), it is apparent that (type 2) diesel properties are close to 5W-30 motor oil properties for engine conditions, except for dynamic viscosity, which his higher for the engine oil but still in a similar magnitude as diesel when compared to air. To have well defined material properties for the liquid phase, measured material properties for the 5W-30 engine oil at 60°C as described in [41] are used for the rest of this work.

## Multiphase Flow

The two-phase flow is described by the following characteristic numbers: The Weber number $We$ relates inertia to surface tension forces and is used to characterize the cohesion or atomization of droplets. It is given as

$$We = \frac{\rho u^2 L}{\gamma},$$

with density $\rho$, relative velocity $u$, characteristic length $L$ and surface tension $\gamma$. The Bond number $Bo$ (also called Eötvös number) describes the ratio of acceleration field forces (in our use case from the relative acceleration of the frame of reference, not earth's gravity) to surface tension forces and is defined as

$$Bo = \frac{\Delta \rho g L^2}{\gamma},$$

with density difference $\Delta \rho$, and acceleration magnitude $g$. Based on the characteristic numbers in Tab. 4, both Weber and Bond number are in an intermediate range. In combination with the Reynolds numbers observed in the blow-by gas analysis, we cannot neglect a-priori any effect of surface tension, acceleration, and viscous behavior. This means, that, in contrast to (gas-only) blow-by calculation, the acceleration force must also be considered for the whole fluid domain. In this work, the level-set method is used to model the multiphase flow.

| property | symbol | value or range | unit |
|---|---|---|---|
| density of diesel | $\rho_f$ | 875–959 | kg/m³ |
| density of air @ 400K | $\rho_a$ | 0.5 | kg/m³ |
| characteristic velocity | $v$ | 80 | m/s |
| characteristic length | $L$ | $1.0 \times 10^{-6}$ | m |
| surface tension | $\gamma$ | $28.73 \times 10^{-3}$ [41] | N/m |
| Weber number | $We$ | $\approx 200$ | [−] |
| density difference | $\Delta \rho$ | $\approx 900$ | kg/m³ |
| peak acceleration magnitude | $g$ | $\approx 1427$ | m/s² |
| Bond number | $Bo$ | $\approx 0.1$ | [−] |

Table 4: Two-phase flow parameters.

## Stabilized Space-Time Finite Element Method

The space-time discretization method [18] is used to discretize the compressible flow, elastic fluid domain movement, and level-set transport problems. Thus, space and time are discretized at the same time using prismatic-in-time and discontinuous-in-time first-order finite elements, which form a so-called space-time slab. Residual-based stabilization terms are added to balance numerical diffusion in a consistent way. Fig. 6 displays an example of a 2D domain where the time is displayed as third dimension. This discretization handles moving domain boundaries intuitively.

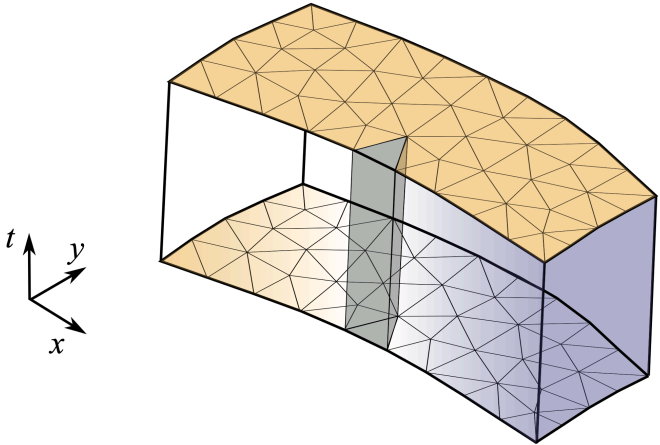

Figure 6 Space-time discretization.

# Transient Blow-by Calculations with Fixed Piston Rings

## Boundary Conditions

In this transient simulation, the piston rings are held fixed while the inflow pressure and temperature are set to a profile representing a full combustion cycle. This step is meant to investigate the transient flow behavior of gas in the piston ring pack.

The transient inflow pressure distribution is taken from measured combustion chamber pressures with piezo pressure sensors at constant test-rig conditions with 1500 RPM and 28 Nm (see Fig. 7).

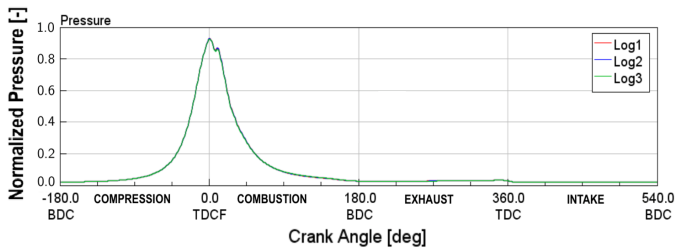

Figure 7: Measured combustion chamber pressure of three different measurements.

A transient temperature measurement within the combustion chamber is hard to realize. Therefore, the inflow (see Fig. 3c) temperature at the piston outer crown land has been calculated with a calibrated 1D GT-Power model. The transient result of the time dependent but spatially constant temperature distribution at the gap between the piston crown and liner is plotted in Fig. 8.

The outflow (see Fig. 3c) temperature is assumed to be constant and set to the measured temperature of the oil in the oil pan with approx. 80°C = 353.15K. The air pressure within the crank case has been measured as $p_{out} = 1.0171 \times 10^5$Pa.

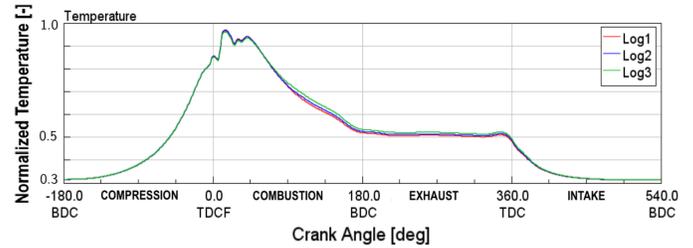

Figure 8: Combustion chamber temperature of three different measurements (re-engineered by 1D simulation).

## Results for Fixed Piston Rings

After a mesh convergence study, three configurations are considered to investigate the gas flow in the Ford piston sealing labyrinth without ring movement to see the flow conditions at best sealing and through an opened path:

- (M1-1a) All rings fixed at 2μm below the groove top (all rings at 'closed' upper flank, good sealing)
- (M1-1b) All rings fixed, the top ring at 5μm above the groove bottom, ring 2 and OCR centered ('opened' path, bad sealing)
- (M1-1c) All rings fixed at 2μm above the groove bottom (all rings at 'closed' lower flank, good sealing)

For the 'closed' cases, a residual gap of 2μm remains between the flank and ring as described above.

The initial condition distorts the transient solution at the beginning of the simulation, as is visible in the initial peak in blow-by (see −180°CA to −100°CA in Fig. 9). Compared to the initial condition from the steady-state investigations, a more appropriate progression is the computed solution of the second cycle. This is why the simulation is continued for 150°CA after reaching 540°CA to get more meaningful values for the beginning of the compression stroke.

The resulting transient pressure distributions along the ring pack is visualized for different degrees of crank angle in Fig. 10 for load case M1-1c. The near ambient pressure near the crank case (bottom part of the figure) shows a good sealing performance of the piston rings during the full cycle. The combustion pressure profile reaches the upper and inner (left in each sub-figure) part of the first sealing ring without significant delay due to the fully opened cross-sections. The leakage and blow-by along the first and second piston sealing rings at the lower flanks cause a pressure increase in the land below, respectively. Compared to the absolute pressure changes during and after combustion, the expansion and intake stroke show only minor changes in the pressure field (not shown in Fig. 10).
Fig. 11 shows the blow-by mass flux for the three variants with fixed rings. Notably, the variant with bad sealing shows a significantly higher blow-by than the variants with piston rings sealing either at the top or bottom groove. A case with all rings centered is not presented, as it was challenging to compute due to the high fluid velocities. We assume that the blow-by peaks would be even higher in that configuration, because our 'bad sealing' case shows this tendency. The two 'closed' crank angle (ITDC at $0°$) variants (best sealing) differ only slightly from each other, as expected due to the similar smallest cross-section of the flow path of the gas. The timespan with the highest simulated blow-by ($0°$CA to $100°$CA in Fig. 11) aligns with the combustion chamber pressure peak ($-45°$CA to $45°$CA in Fig. 7) at the ignition with a delay.

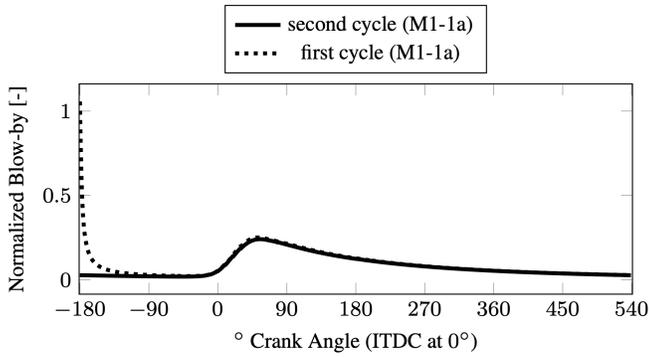

Figure 9: Influence of initial condition on blow-by at outflow over one work cycle.

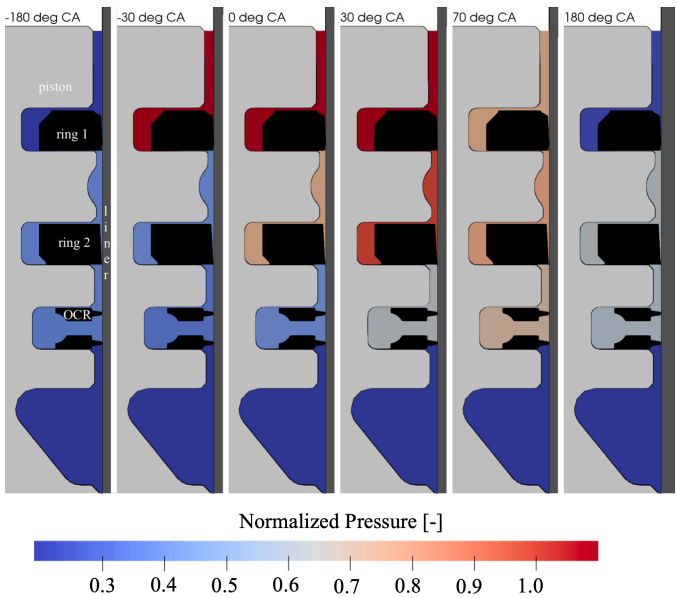

Figure 10: Pressure field for fixed piston rings all at lower flank (M1-1c, good sealing load case).

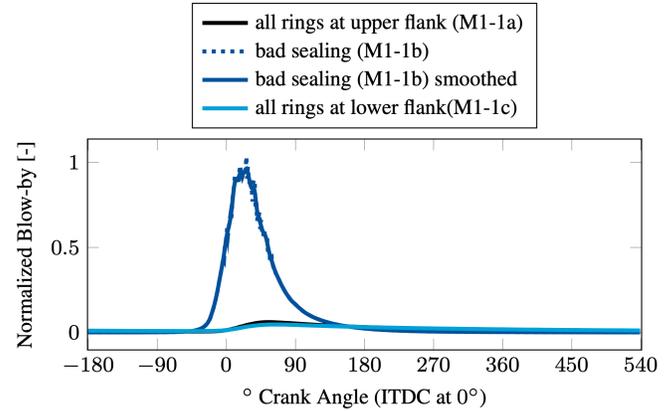

Figure 11: Blow-by at outflow over one work cycle for fixed ring configurations.

## Moving Piston Rings and Influence of Ring Motion on Gas Blow-by

To investigate a more realistic gas flow in the piston sealing labyrinth using moving piston rings, the 2D-test case from the previous section has been extended by a model for moving domain boundaries.

### Modeling of Ring Movement

The position of the piston rings inside their respective grooves characterizes the flow path of blow-by gases, fuel, and oil from the combustion chamber to the crankcase through the sealing labyrinth of the piston rings. Only the forces on the piston rings from the relative piston movement are known, but not the friction forces of the oil-lubricated ring-liner contact. Thus, the piston movement profiles of a similar engine are taken from literature [16].
As described in [16], the oil control ring has more contact surface to the cylinder and thus more friction and moves in the groove with slower speeds at BDC and TDC than at other piston positions in the cylinder.

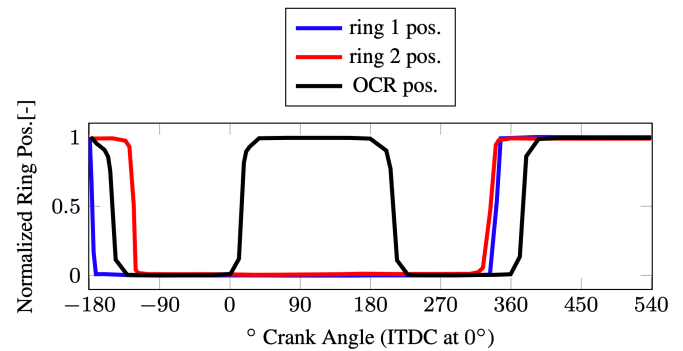

Figure 12: Used piston ring movement profile with normalized ring position and combustion chamber pressure over crank angle (ITDC at $0°$).

The model of the piston ring movement in axial direction consists of flank changes from the upper to lower position and vice versa. During the combustion stroke ($0°$CA to $180°$CA), the high combustion pressure is permanently acting on the first and second piston ring, pressing them firmly to their respective lower flanks (see Fig. 12).

Within this work, neither radial collapse of the piston rings nor variation of the cylinder liner wall is considered.
The ring motion profile suggests that the oil control ring (OCR) moves up during combustion and after the exhaust stroke (see Fig. 12). Additional flutter movements (incomplete flank changes) are measured for the first sealing ring. The flank changes occur less instantaneous for the lower rings. The OCR shows flank changes with a slower start and end of movement near the upper and lower flank, probably due to damping by more friction between the piston ring and cylinder liner [9]. Since a slow flank change leaves the leaking path through the groove open for a longer time than a faster flank change, we expect that flank change timing significantly influences the blow-by and possible fuel or oil transported through the piston ring grooves by the blow-by. Besides better sealing and smaller piston ring gap leakage, a different closing and opening timing of the piston rings may improve the sealing behavior. Therefore, this work investigates different timings of piston ring flank changes.

### *Transient Blow-By Calculations with Moving Piston Rings*

A residual gap buffer zone around the groove walls is left open, so that the smallest possible gap between the groove flank and piston ring is approx. 2μm on the upper and lower groove flank for all rings. Furthermore, the gap between the outer ring wall and liner wall is left open at 1μm.

### *Results with Prescribed Ring Movement*

The ring profiles of Oliva et al. [16] were obtained for a similar engine but without a crank offset. Thus, several variations in piston ring flank change timing are included in this work's analysis:

- (M1-2a) Ring movement as in Fig. 12
- (M1-2b) M1-2a, but 3°CA earlier
- (M1-2c) M1-2a, but with a 10% flank change speed increase

The gas pressure field in the ring pack of the M1-2a case, presented in Fig. 13, shows higher second land pressures during combustion compared to the case with fixed sealing rings. The corresponding temperature field is displayed in Fig. 14, and the velocity magnitude field in Fig. 15. A significant blow-by flow into the second and third land can be observed at 0°CA and after during the combustion phase. The subsequent flank change of the OCR leads to gas leaking into the crankcase (see 16°CA, 38°CA, and 160°CA plots in Fig. 15). An additional blow-by passing the first piston ring is visible during its flank change at 340°CA and past the OCR at its respective flank change at 375°CA. During the flank changes, the Mach number exceeds the nearly incompressible regime (see Fig. 16). With Mach numbers at or above 1, some sections of the domain contain shocks and compressibility effects, which could not have been observed with incompressible flow models. This confirms the initial assumption of compressible flow physics.

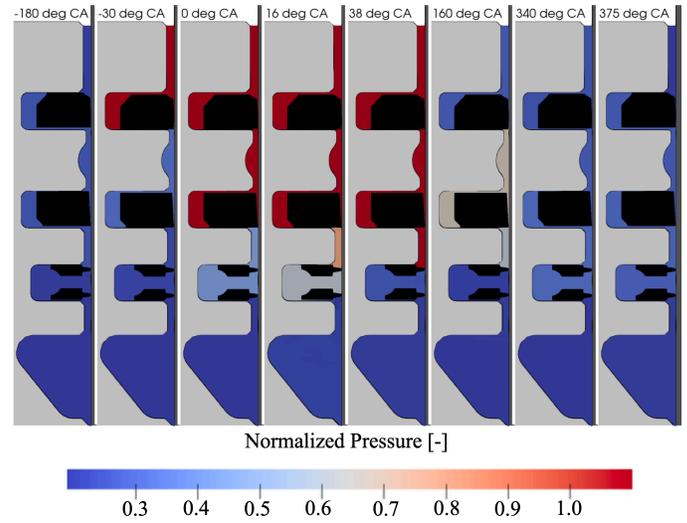

Figure 13: Pressure with moving piston rings (M1-2a).

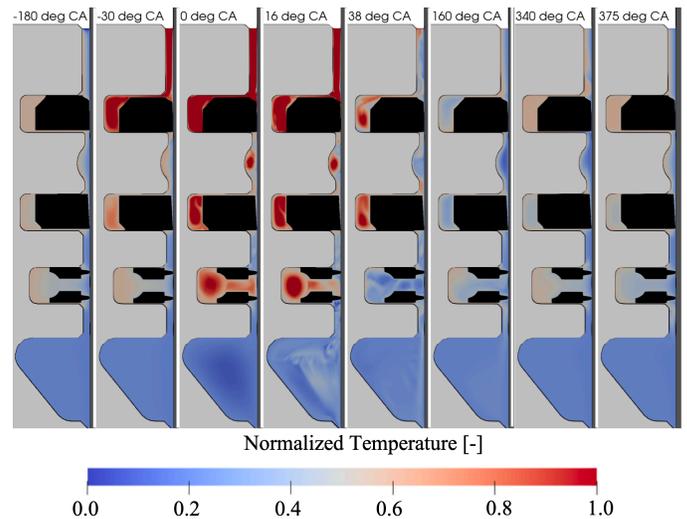

Figure 14: Temperature with moving piston rings (M1-2a).

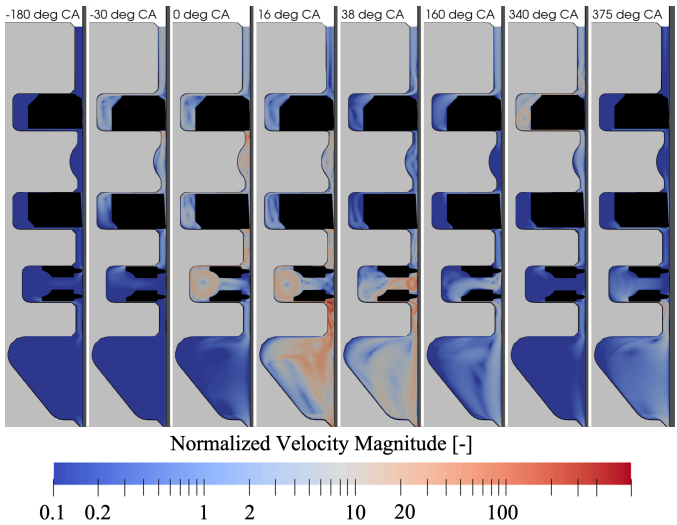

Figure 15: Velocity magnitude with moving piston rings (M1-2a).

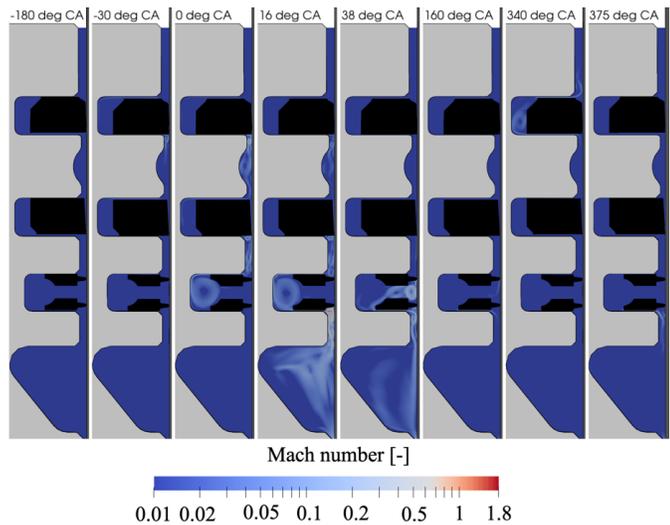

Figure 16: Mach number with moving piston rings (M1-2a).

The base run with moving piston rings (M1-2a) also has the highest peak in blow-by after the maximum inflow pressure at ITDC (presented in Fig. 17d). This occurs at 16°CA − 18°CA. Additional smaller spikes of blow-by can be observed when the OCR moves at 205°CA and 375°CA (see Fig. 17d). The first flank change before combustion results in no significant spike in blow-by. The predicted mean blow-by value of run M1-2a is in the range of the measured values of blow-by experiments and hits the absolute level well.

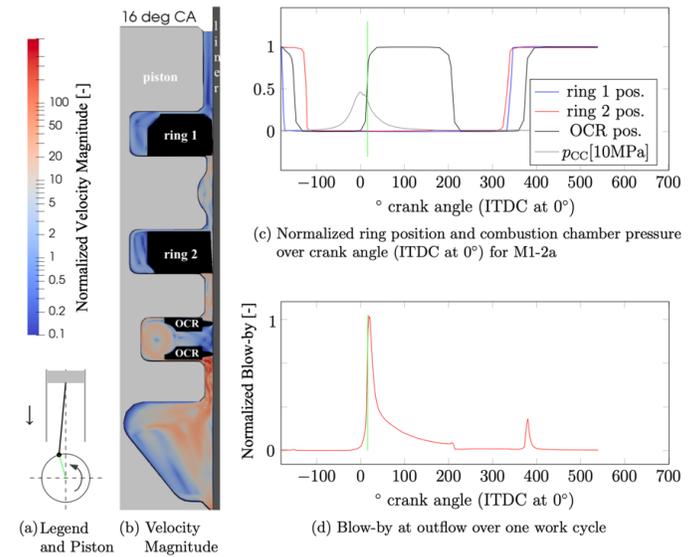

Figure 17: Piston position inside cylinder at 16°CA, qualitative velocity magnitude (red high and blue low) inside piston ring pack, ring movement and blow-by over full cycle for M1-2a case.

The variation with an earlier ring movement (M1-2b, see Fig. 18) shows a similar blow-by profile as the base moving rings version, but with earlier peaks in the blow-by after TDC. The blow-by peak is barely lower than it was for the later ring movement (M1-2a).

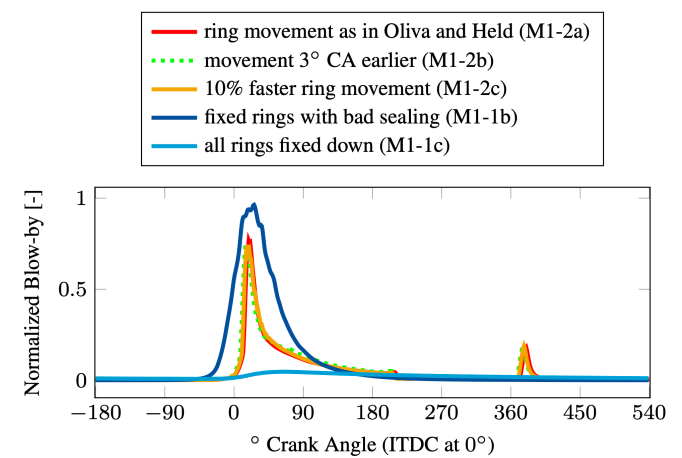

Figure 18: Blow-by at outflow over one work cycle.

The resulting blow-by profile for faster moving rings (M1-2c, orange in Fig. 18) is like the profile with earlier moving rings (M1-2b). Additionally, the peak in blow-by mass flux after ITDC is lower compared to the other cases with moving piston rings (M1-2a/b). For both variation cases M1-2b/c at the smaller peak around 375°CA, the blow-by mass flux is approx. 5% higher than for the reference case with moving rings (M1-2a). The mean blow-by values are increased for M1-2b (+10%) and M1-2c (+4%) compared to M1-2a. As expected, the blow-by values of all variants with moving rings are bounded from above by the bad sealing case and bounded from below by the best sealing case (see fixed piston rings) for most of the combustion phase.

## Two-Phase Flow Results

This section includes application cases for flow phenomena in different parts of the piston ring pack. The flow is modeled using the material model for two non-mixing fluids as described above. Material properties for the 5W-30 oil at 60°C are used for the liquid phase, even for the fuel flow. Thus, 'fuel' and 'oil' both refer to the liquid phase and are used interchangeably in this chapter. The meshes for these cases were generated with triangular first order elements.

### *Groove with Sloshing Fuel or Oil*

This case is an example for the piston ring pack flow in the piston ring grooves. A fuel or oil droplet is moving in an air-filled groove behind a piston ring while the piston follows the engine crankshaft movement at 1500 RPM. The simulated geometry is rectangular with dimensions like a groove cavity behind a piston ring. In contrast to the mesh for the piston groove region, the mesh for this test case has uniform element sizes. The acceleration experienced by the fluids in the moving reference system is applied as a volume term proportional but opposite to the acceleration of the piston ($g = -\hat{j}g\cos(\omega t)$, where $\hat{j}$ is the unit vector along the piston stroke direction).

With the Bond and Reynolds number limited by the values in Tab. 5, the test case transitions between spherical and ellipsoidal droplet regimes as described in [42]. The used time step size is $2 \times 10^{-6}$s.

| | | |
|---|---|---|
| $\rho_1$ | density of air @ 400K | 0.5kg/m³ |
| $\rho_2$ | density of droplet | 809.5kg/m³ |
| $g$ | peak volume acceleration term | 1106m/s² |
| PGH | piston groove height | 2.0mm |
| $h$ | mesh element length | 12.5µm |
| $d$ | droplet diameter | $0.5 \times 10^{-3}$m |
| $\mu_2$ | dynamic viscosity of droplet | 0.00873kg/ms |
| $\gamma$ | surface tension coefficient | 0.03N/m |
| Re | Reynolds number | $\frac{\rho_2\sqrt{gd}d}{\mu_2} \cong 12$ |
| Bo | Bond number | $\frac{\rho_2 d^2 g}{\gamma} \cong 1.8$ |

Table 5: Parameters within the groove of fuel or oil test case.

The fuel droplet is positioned initially in the lower half of the groove (see Fig. 19). It is accelerated first against the lower wall of the groove-like cavity, then upward toward the top part. No sticking behavior to the wall is observed with this model. Note that, in this case, the surface tension forces stop the droplet from dispersing into smaller droplets.

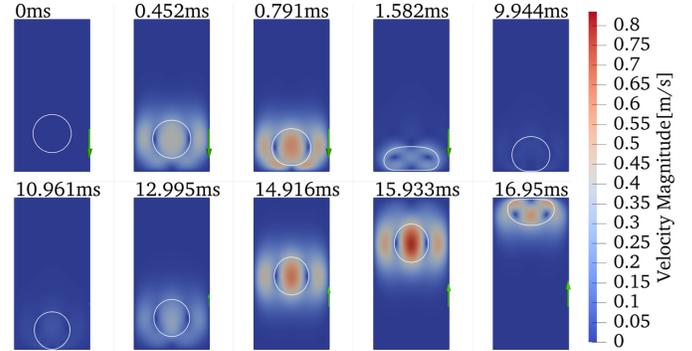

Figure 19: Velocity field for one droplet in a piston ring groove with interface in white and normalized acceleration as green arrow.

### *Fuel Droplet Flow Between Piston and Fixed Piston Ring*

In this case, the area around the bottom chamfer of the oil control ring (OCR) is investigated. The computational domain extends from the lower inner edge of the OCR down to the bottom chamfer of the land below the OCR, as well as up to the bottom contact point of the OCR and cylinder liner or block wall (see Fig. 20). For this investigation, we assume this contact point (top right in Fig. 20) to be closed off without a residual gap. The geometry and locations of the inlet and outlet are documented in Fig. 20. In this case, the OCR is fixed, so the upper wall does not move.

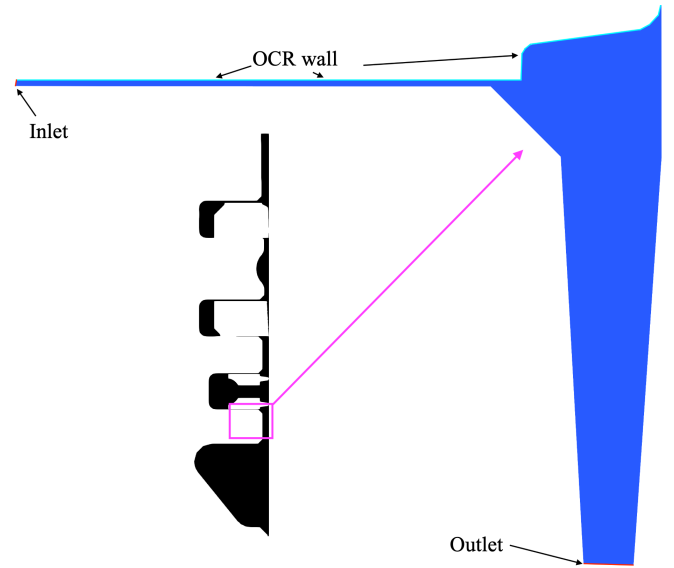

Figure 20: Fluid domain geometry (blue) of the OCR chamfer test case with inlet and outlet in red and OCR wall in light blue (moving unless noted otherwise).

The initial velocities are assumed to be 0m/s and the pressure is uniformly initialized to the crankcase pressure. The temperature field is initialized with 400K.

The inlet pressure is prescribed as values extracted from the same location from the investigation of the single phase flow of gas and moving piston rings for the load case of 1500 RPM and 28 Nm (case M1-2a), starting at the equivalent time of 17°CA before ITDC (18.11ms after BDC), see also Fig. 21.

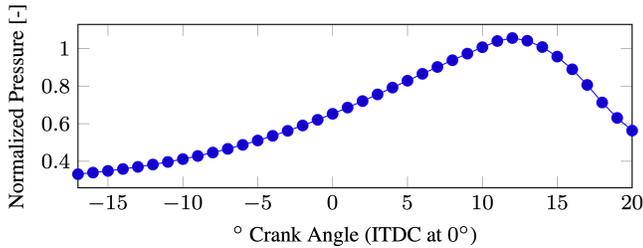

Figure 21: Pressure at the OCR lower inner edge (inlet).

Since temperature variations are assumed to be minor on the walls of the geometry, the temperature is prescribed to be 400K on all walls and the inlet, but not at the outlet. No-slip boundary conditions are enforced strongly on all walls. Volume forces are accounted for as source terms in the momentum equation like in the previous section, see Fig. 22. The piston acceleration is negative in the simulated time interval, so the volume acceleration term is equal in magnitude but with an opposite sign to account for the moving reference frame.

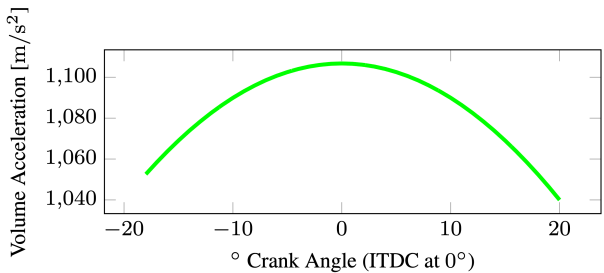

Figure 22: Volume acceleration $g$ component in piston primary movement direction.

Since the mesh elements are smaller ($h_{min} \approx 0.02\mu m$) than in the gas-only test case in the same region, the time step is reduced to $5.0 \times 10^{-7}$s. A section of the groove between the lower piston flank and oil control ring, including the inlet, is initially filled with liquid (fuel or oil). The rest of the domain is initially filled with gas.

The fuel is pushed out of the groove toward the cylinder liner, see Fig. 23b. As soon as the fuel has left the chamfer far enough to leave a path for the gas toward the outlet around the fuel, the gas velocity rapidly increases and drags the droplet toward the liner (see Fig. 23c). Smaller droplets of fuel stick to the OCR and piston chamfer (see Fig. 23d). The main fuel fragment moves horizontally, sticks to the cylinder liner upon reaching it, and moves slowly toward the outlet for the rest of the simulation (see Fig. 23h).

From a numerical point of view, the global mass correction scheme for the level-set method can keep the total amount of fuel within the domain constant, but the individual droplet masses are not conserved. With increasing simulation time, the error in the mass distribution increases as mass is moved from the chamfer droplet to the fuel fragment sticking to the OCR (compare Figs. 23g vs. 23h). This effect can be reduced with a finer resolution in space and time but cannot be eliminated for the classical level-set approach.

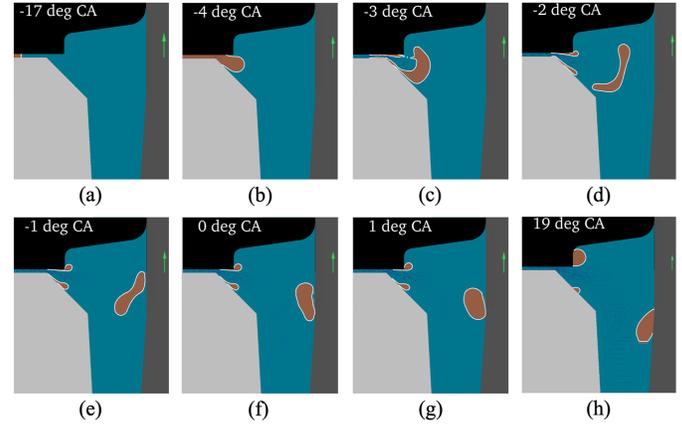

Figure 23: Fuel droplet leaking through the gap between the fixed OCR (black) and piston bottom groove with chamfer (light gray) and hitting the cylinder liner (dark gray). The fluid density is displayed from blue (air) to orange (fuel or oil) with white interface contours. The normalized acceleration is shown as green arrow.

### Fuel Flow Between Piston and Moving Piston Ring

For this scenario, the previous test case is modified in the following way: The top wall of the domain (light blue in Fig. 20) is displaced according to the movement of the OCR, see Fig. 24. The moving fluid domain is tracked with the EMUM. Additionally, the inlet species is set to fuel instead of gas.

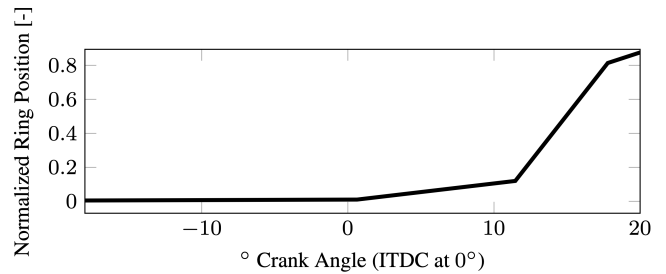

Figure 24: OCR position over crank angle.

For crank angles up until 10°CA after ITDC, the fuel behavior is like the previous case. However, as soon as the OCR moves up significantly, the flow velocities in the gap increase (see Fig. 25c). The fuel droplet forming in the chamfer is pushed toward the cylinder liner by a fuel jet out of the gap, see Fig. 25d. This fuel jet hits the liner and is transported along it toward the outlet (see Fig. 25e). Ligaments and droplets detach from the jet, but it does not atomize. The liquid Reynolds number is at approx. 50 with the liquid Weber number at approx. 100. The gas Weber number is at approx. 0.026. In the classification from [43], this case is in the category where a jet appears ($We_L \approx 100 > 8$) and is in the first wind regime with $We_G <$ 0.4. As a result, the ambient gas in no longer negligible. The simulated jet hits the liner without breaking up. This corresponds well with the linear theory analysis from [43], which predicts a jet breakup only after a distance of several times the gap width between the piston ring and groove.

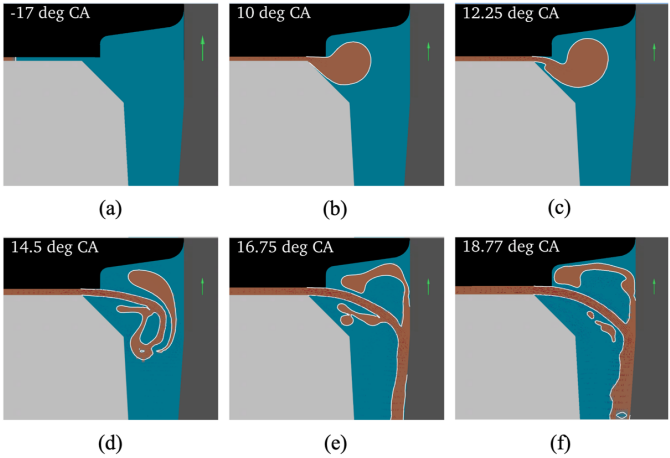

Figure 25: Fuel jet leaking through the groove between the moving OCR (black) and piston bottom groove with chamfer (light gray) and hitting the cylinder liner (dark gray). The fluid density is displayed from blue (air) to orange (fuel or oil) with white interface contours. The normalized volume acceleration is shown as green arrow.

## Inlet Fuel Film Entering the Crown Land Region

The final application case mimics the late post-injection act. The focus is on the transport of the liquid film swept into the crown land and upper side of the first piston sealing ring via the upward motion of the piston at the end of the exhaust stroke. The engine load parameters are again chosen for a load case of 1500 RPM and 28 Nm and material values of 5W-30 engine oil at 60°C are used for the liquid phase.

### Inlet Film

The film on the cylinder liner or block wall consists of an oil or fuel mixture with a thickness of up to 10μm [44]. At the spray target of the fuel injector jets, the film mixes with fuel parts of the late post-injection, starting later than 130-140°CA after ITDC with less than 7mg fuel per cycle for this operation point. The position and fuel distribution of the late post-injection residue is modeled using the following simplifications:

In this investigation, the spray target is approximately 11mm wide in piston stroke direction, starting at its topmost point at 4mm and ending at 15mm below the injector. The piston is far below this region for the second half of the combustion stroke so that unburned fuel fragments hit the liner walls. The piston only encounters the fuel in the spray target during the (upward) exhaust stroke at 320°CA to 340°CA (with ≈ 6m/s to 3m/s) for the thickest part, which corresponds to $\frac{510}{9000}$ s $- \frac{530}{9000}$ s after compression BDC. There is a slight variation in the spray target thickness since the lower part disperses into a larger area than the upper part. Since the target film might be thicker than just the deposited fuel due to unburned fuel or oil components on the cylinder wall, we assume a constant incoming base film thickness of 10μm when the piston sweeps along the fuel film.

The computational domain for this case, presented in Fig. 26, consists of the crown land, the top chamfer at the first piston sealing ring, the gap between first sealing ring top flank and its groove, as well as the cavity between the first piston sealing ring's outer side and the cylinder liner up to the point where the first piston ring (ring 1) and liner are closest. As ring 1 moves, the fluid domain deforms over time. This deformation is tracked using EMUM; the only non-trivial boundary conditions for the mesh motion problem are given by the expression for the motion of ring 1.

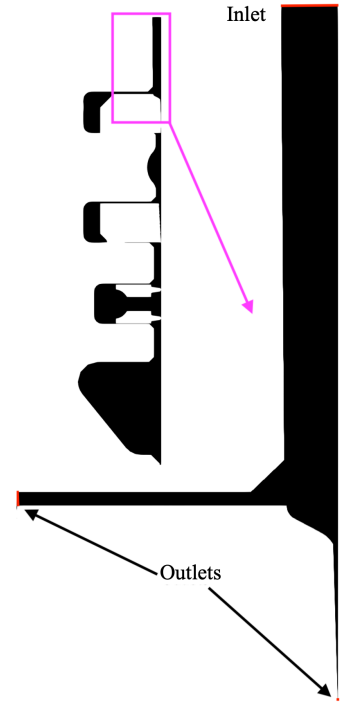

Figure 26: Geometry of the inlet film case with inlet and outlets in red.

Compared to the single-species (gas) simulations with 20,260 elements in the full domain, the mesh for this case consists of 128,112 smaller elements for just the piston chamfer region. For a direct comparison, see Fig. 27.

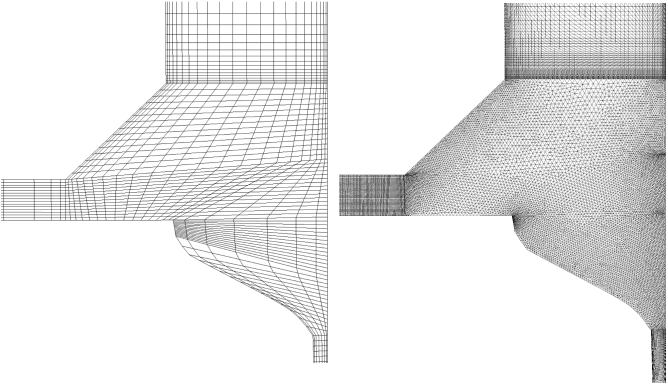

Figure 27: Close-up on the computational meshes of the piston chamfer region for both the gas-only simulations (left) and the two-phase flow in the inlet region (right).

Time-dependent pressure boundary conditions for the three open boundaries are used from the M1-2 case. The inlet film is represented by a Dirichlet boundary condition for the level-set field signed distance function on the combustion chamber inlet.

$$\phi_{\text{inlet}} = -x + 10^{-4}\left(1 - H\left(-\left(\left|t - \frac{510}{9000}\right| - \frac{10}{9000}\right)\right)\right),$$

where H is the Heaviside function and $x$ the coordinate in radial direction (with the origin on the entering point of the film and the coordinate of the liner wall at $x_{liner} = 10^{-5}$m).

This expression sets the inlet phase to air for most of the simulation (0μm inlet fuel film thickness) but sets the inlet fuel film thickness to 10μm for the time interval where the piston top sweeps along the spray target.

To account for the relative motion, the velocity on the liner wall is set to the time-dependent velocity of the piston with an opposite sign. The remaining parameters for the two-phase flow are set like in the previous test case.

## *Results for Inlet Film at Spray Target*

The simulated flow field (see Fig. 28) shows few differences compared to the simulated gas flow field from case M1-2 (see section above), namely: While the piston assembly is moving up, the fuel film entering the flow domain along the cylinder wall after 320°CA (see Fig. 28b) accumulates to a thick tip that splashes between ring 1 and the liner (see Fig. 28f). The fuel splashes back up and toward the gap between ring 1 and the piston chamfer above it (see Fig. 28g). At about 335°CA, the (gas) flow through the top piston ring flank gap (over ring 1) reverses and flows back toward the combustion chamber (reverse blow-by), dragging the fuel droplets with it into the chamfer region (see Fig. 28h). Thus, this simulation configuration with all its assumptions does not predict the fuel scraped up from the late post-injection entering the piston ring labyrinth.

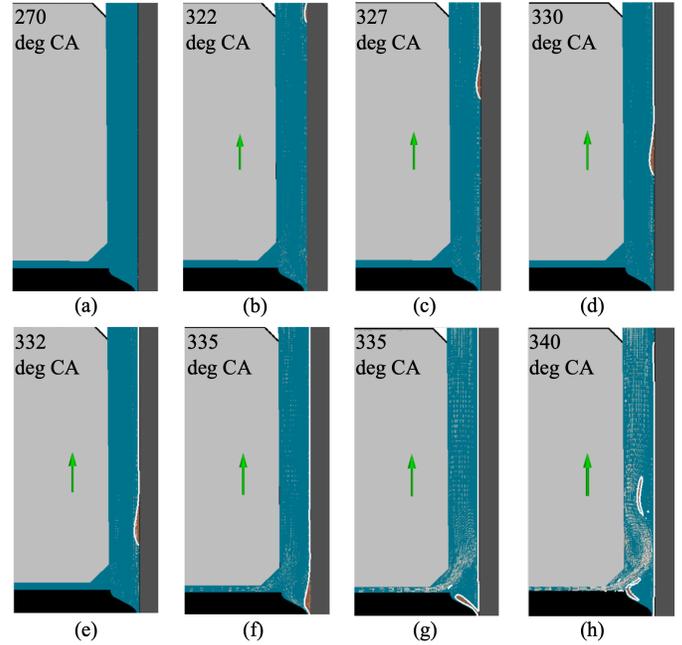

Figure 28: Fuel entering the inlet of the crown land with piston in light gray, cylinder liner in dark gray, piston ring in black, fuel and air interface as white contour, volume acceleration direction as green arrow, fluid density ranging from blue (air) to orange (fuel), and velocity as white glyphs.

Variants of the simulation with thicker inlet fuel films (2 and 3 times thicker) show similar behavior due to the change from stagnation to a reverse blow-by starting too early to permit fuel entering the piston ring gap (see Fig. 29).

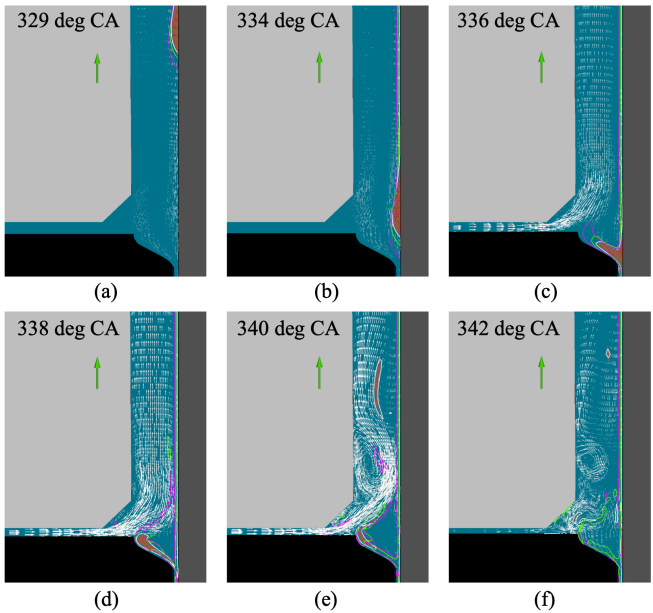

Figure 29: Closeup on the fuel film and air flow in the piston chamfer with piston in light gray, cylinder liner in dark gray, first piston sealing ring in black, volume acceleration direction as green arrow, density field in blue (air) to orange (fuel), and velocity as white glyphs. The fuel and air interface is represented as a contour for the film inlet thicknesses of 10μm (white), 20μm (green), and 30μm (pink).

## Inlet Film from Fuel Early in the Exhaust Stroke

Since the early reversal of the flow direction in the top piston ring gap rules out a fuel transport toward the groove cavity with the current model configuration (even with exaggerated film thicknesses), other scenarios for a fuel transport toward the crankcase should be explored: One scenario could be a ring-liner collapse, where the blow-by gases are transporting fuel directly into the second land (not investigated in this work). Or fuel is transported more downward along the cylinder and is scraped up at a lower position in the cylinder (earlier in the exhaust stroke).

The latter scenario requires enough fuel on the cylinder liner far below the spray target. In a real engine, the unburned fuel from the late post-injection is concentrated at the liner target of the injector jets and below. Due to the high injector pressure of more than 2000bar, the unburned fuel may travel downward along the liner wall toward the first piston ring. This fuel below the target would be encountered by the piston earlier in the exhaust stroke compared to the fuel at the injector target.

To analyze this case, the simulation is started at 200°CA after ITDC, with a thick film of fuel on the cylinder liner encountered by the moving frame of reference for 20°CA. In contrast to the previous case, the computational domain now includes the full ring liner clearance up until the lower chamfer of the first piston ring. Also, because the first sealing ring does not move significantly until 344°CA after ITDC, the mesh is static.

The inlet film fills the cavity between the first piston ring and liner (see Fig. 30c).

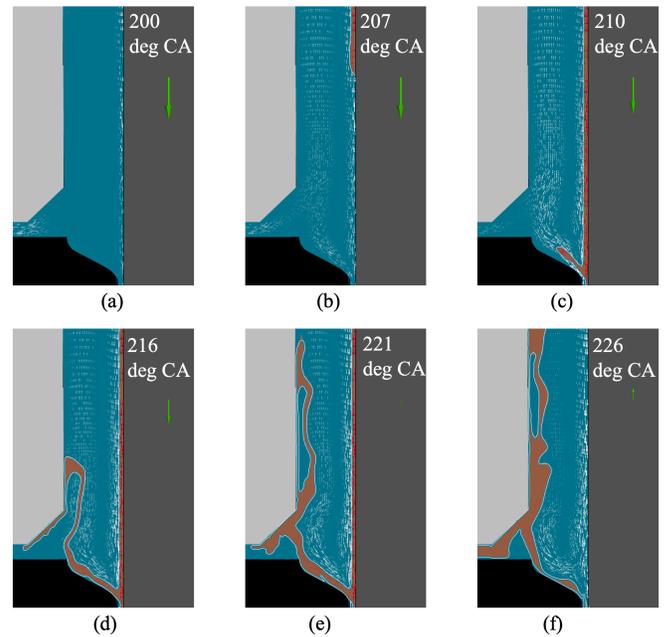

Figure 30: Closeup on the fuel transport in the piston chamfer region for an early inlet film case with piston in light gray, cylinder liner in dark gray, first piston sealing ring in black, volume acceleration direction as green arrow, density field in blue (air) to orange (fuel), and velocity as white glyphs.

Most of the inlet film is redirected by the contour of the first piston ring toward the crown land and chamfer at the top of the ring gap (see Fig. 30d). Even though a large part of the incoming fuel is redirected toward the crown land, some fuel enters the upper ring gap of the first piston ring (see Figs. 30e and 30f). The redirected fuel flow hits the 45° piston chamfer and splits into two paths at the stagnation point: One path runs toward the upper crown land and the other path toward the piston ring gap (see Fig. 30f). With a period of more than 100°CA of piston movement remaining until the (gas) flow in the gap reverses and the upper flank is closed off by the piston ring, it is likely that some leaking fluid enters the groove behind the piston ring and remains there until the next engine cycle. From an engineering point of view, the second fuel flow path toward the piston ring gap should be avoided or heavily decreased to reduce fuel-in-oil leakage.

## Design Proposals

Since the redirection of the inlet film is caused by the geometry of the first piston-ring upper contour and the piston chamfer above it, a re-design of these parts could reduce the amount of fuel entering the piston ring pack. The following design proposals should be considered to adjust the chamfer relative to the flow's stagnation point to move the stagnation point of the incoming film nearly out of or above the chamfer:

- Reducing the piston chamfer size (see Fig. 31b).
- Adjusting the piston chamfer angle and thus its height (see Fig. 31c).

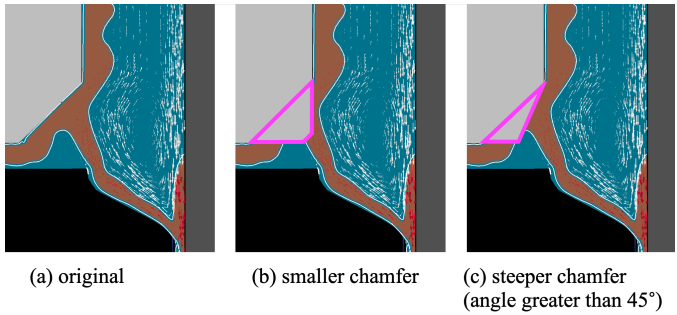

(a) original    (b) smaller chamfer    (c) steeper chamfer (angle greater than 45°)

Figure 31: Chamfer modification. Fuel transport at the piston chamfer and piston ring geometry with density field from blue (air) to orange (fuel) for the analyzed (unmodified) case and piston chamfer design variations (highlighted in magenta) superimposed on the original flow field.

Further design proposals include the re-shaping of the piston-ring top contour to turn the flow earlier such that it hits the piston higher above the piston chamfer:

- Moving the curved part of the piston ring top contour toward the liner (horizontal about distance $d_{curve}$ in the plots to the right side, see Fig. 32b).
- Increasing the ramp angle of the piston ring contour $\beta$ (see Fig. 32c).

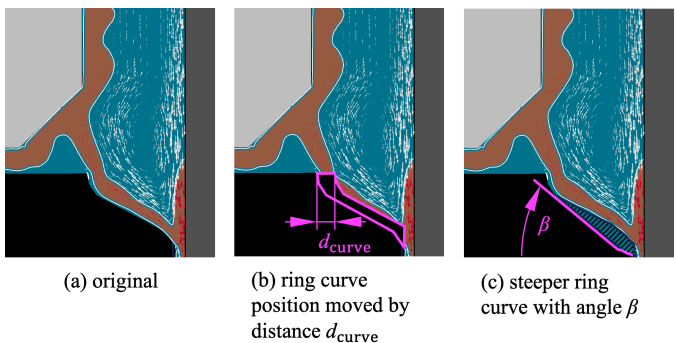

(a) original    (b) ring curve position moved by distance $d_{curve}$    (c) steeper ring curve with angle $\beta$

Figure 32: Re-shaping of the piston ring top contour. Fuel transport at the piston chamfer and piston ring geometry with density field from blue (air) to orange (fuel) for the analyzed (unmodified) case and piston ring design variations (highlighted in magenta) superimposed on the original flow field.

Combining the design proposals described above using smaller geometrical modifications of each single action may result in a beneficial fuel flow not massively entering the first piston gap and groove.

## Summary and Conclusion

This work has investigated the blow-by gas and the gas and fuel two-phase flow around the piston ring pack of a Ford 2.0L I4 diesel test engine using CFD simulations. The computational model uses a compressible fluid formulation as direct flow simulation and includes the domain movement of the piston rings. Expressions for the piston primary movement were re-derived for crank and pin offset to be used in boundary conditions and volume acceleration terms.

Simulations were conducted with boundary data from experiments and 1D-simulations.

Transient blow-by simulations with fixed piston rings confirmed a significant influence of the piston ring position within the piston ring grooves on blow-by as well as computing cost. Vertically centered ("floating") piston rings had the worst sealing performance and the high blow-by velocities led to increased computational costs due to small critical time steps and thus, a large number of computed time steps.

Results of transient blow-by simulations with piston ring movement profiles from literature showed a significant influence on the blow-by mass flux of the piston-ring flank change starting time and duration. The computed flow field corresponded to transonic and supersonic Mach numbers close to the chamfers during piston ring flank changes. This confirms the initially hypothesized compressible behavior of blow-by gases in the piston ring pack. Within this work, the "radial collapse" of the piston rings with a larger gap between the piston ring and liner has not been considered, but it may have some influence on the blow-by.

The used computational model also captures the two-phase flow phenomena of fuel and gas transport using a level-set method with a mass-correction scheme and volumetric surface tension forces. The two-phase flow model proved to be able to capture fuel or oil sloshing in the piston ring groove cavities, as well as liquid droplets and liquid jets in a gas flow through the oil control ring groove bottom region. The liquid jet from the oil control ring groove was identified as laminar and predicted to hit the liner without atomizing for the studied load case.

For a late post-injection, a simplified model for the fuel film thickness on the liner has been developed. The transport of the fuel film into the piston ring pack was studied for different fuel film thicknesses and moving piston rings, but without a ring collapse. In the case where the piston encounters a liquid film near the spray target, no liquid from the film was transported toward the piston-ring groove. In the case of a liquid film being encountered early in the exhaust stroke, some liquid entered the upper flank of the first piston ring. To reduce fuel leakage from the scraped-up inlet film into the first piston-ring gap, design directions have been proposed to modify the piston chamfer above the first sealing ring and adjust the first piston-ring top contour to redirect the fuel film toward the crown land instead of flowing into the piston-ring gap.

Overall, the analyses demonstrate that modern CFD tools like the level-set method are a powerful way to investigate multiphase flows through the piston ring pack to provide significant insights into the processes involved and to reduce the fuel-in-oil leakage within internal combustion engines.

## Acknowledgements

This work was supported by the Alliance Program between Ford Motor Company and RWTH Aachen University.
The authors gratefully acknowledge the computing time granted by the JARA Vergabegremium and provided on the JARA Partition part of the supercomputer CLAIX at RWTH Aachen University.
The authors gratefully acknowledge the computing time provided to them on the high-performance computer Lichtenberg at the NHR

Centers NHR4CES at TU Darmstadt. This is funded by the Federal Ministry of Education and Research, and the state governments participating on the basis of the resolutions of the GWK for national high performance computing at universities (www.nhr-verein.de/unsere-partner).
Funded by the Deutsche Forschungsgemeinschaft (DFG, German Research Foundation) – 537928890; and under Germany´s Excellence Strategy – EXC-2023 – 390621612.

## Definitions/Abbreviations

| | |
|---|---|
| **BDC** | Bottom Dead Center |
| **CA** | Crank Angle |
| **CAD** | Computer-Aided Design |
| **CFD** | Computational Fluid Dynamics |
| **CLSVOF** | Coupled Level Set Volume Of Fluid |
| **e-fuel** | electrofuel or synthetic fuel |
| **EMUM** | Elastic Mesh Update Method |
| **FEM** | Finite Element Method |
| **FVM** | Finite Volume Method |
| **GMRES** | Generalized Minimal RESidual method |
| **HPC** | High Performance Computing |
| **ICE** | Internal Combustion Engine |
| **ITDC** | Ignition Top Dead Center |
| **MPI** | Message Passing Interface |

| | | | |
|---|---|---|---|
| **OCR** | Oil Control Ring | **TDC** | Top Dead Center |
| **RPM** | Revolutions Per Minute | **VOF** | Volume Of Fluid |